\documentclass[aps,pra,reprint,superscriptaddress]{revtex4-1}
\usepackage{amsmath, amssymb, amsfonts, amsthm}
\usepackage{graphicx}
\usepackage{color}
\usepackage[hidelinks]{hyperref}

\newcommand{\ket}[1]{{\big| {#1} \big>}}
\newcommand{\bra}[1]{{\big< {#1} \big|}}
\newcommand{\expval}[1]{{\big< {#1} \big>}}
\DeclareMathOperator{\tr}{tr}
\newcommand{\irm}{\mathrm{i}}
\newcommand{\drm}{\mathrm{d}}
\newif\ifshowcomments

\showcommentstrue 

\theoremstyle{definition}

\begin{document}

\title{Universal scheme for indirect quantum control}

\author{David Layden}
\affiliation{Department of Applied Mathematics, University of Waterloo, Waterloo, Ontario, N2L 3G1, Canada}
\affiliation{Institute for Quantum Computing, University of Waterloo, Waterloo, Ontario, N2L 3G1, Canada}
\author{Eduardo Mart\'{i}n-Mart\'{i}nez}
\affiliation{Department of Applied Mathematics, University of Waterloo, Waterloo, Ontario, N2L 3G1, Canada}
\affiliation{Institute for Quantum Computing, University of Waterloo, Waterloo, Ontario, N2L 3G1, Canada}
\affiliation{Perimeter Institute for Theoretical Physics, 31 Caroline St N, Waterloo, Ontario, N2L 2Y5, Canada}
\author{Achim Kempf}
\affiliation{Department of Applied Mathematics, University of Waterloo, Waterloo, Ontario, N2L 3G1, Canada}
\affiliation{Institute for Quantum Computing, University of Waterloo, Waterloo, Ontario, N2L 3G1, Canada}
\affiliation{Perimeter Institute for Theoretical Physics, 31 Caroline St N, Waterloo, Ontario, N2L 2Y5, Canada}

\begin{abstract}
We consider a bipartite quantum object, composed of a quantum system and a quantum actuator which is periodically reset. We show that the reduced dynamics of the system approaches unitarity as the reset frequency of the actuator is increased. This phenomenon arises because quantum systems interacting for a short time can impact each other faster than they can become significantly entangled. In the high reset-frequency limit, the effective Hamiltonian describing the system's unitary evolution depends on the state to which the actuator is reset. This makes it possible to indirectly implement a continuous family of effective Hamiltonians on one part of a bipartite quantum object, thereby reducing the problem of indirect control (via a quantum actuator) to the well-studied one of direct quantum control.
\end{abstract}



\maketitle

\section{Introduction}
Coherent control of quantum systems is a central ingredient for most implementations of quantum computing and metrology. In practice, however, interactions between a controlled quantum system and its environment typically lead to a gradual loss of coherence in the system. It is, therefore, desirable to perform control operations rapidly compared to the system's coherence times, so as to minimize environment-induced errors. This goal poses a dilemma,
since by coupling a system more strongly to a classical controller in order to steer it more quickly, the system may also become coupled more strongly to the general environment, which can then increase its rate of decoherence. Conversely, a system with longer coherence times typically interacts more weakly with all of its surroundings, including a classical controller, making it difficult to control rapidly.

An appealing compromise is to employ indirect control with a setup involving two quantum systems: the to-be-controlled quantum system $\mathcal{S}$ which is relatively well isolated, coupled to a quantum actuator $\mathcal{A}$ which is classically controlled and interacts more strongly with the outside world. This strategy has attracted considerable attention since it was found that control of $\mathcal{A}$ can yield universal control of $\mathcal{S}$ \cite{Lloyd2001,Lloyd2004}. Indirect control schemes have shown promise in a variety of settings, including spin chains \cite{Burgath2009, Burgath2010, Heule2010}, superconducting qubits \cite{Strauch2011}, nanomechanical resonators \cite{Jacobs2007,Rabl2009}, and perhaps most notably, in nuclear/electron spin systems \cite{Morton2005,Morton2008,Hodges2008,Cappellaro2009,Taminiau2014,Aiello2015}, where they are commonly used. There is, however, no general recipe for mapping a desired unitary on $\mathcal{S}$ to a series of control operations on $\mathcal{A}$ which will produce it \cite{Jacobs2007}. This is largely due to the entangling nature of the system-actuator coupling; a generic operation on $\mathcal{A}$ leaves it entangled with $\mathcal{S}$, and thus has a net non-unitary effect on the latter.

Physical settings in which a nuclear spin ($\mathcal{S}$) couples to an electron spin ($\mathcal{A}$), such as nitrogen-vacancy (NV) centers and electron spin resonance (ESR) systems, offer an exceptionally convenient way of surmounting this difficulty, as their Hamiltonians can often be cast in the form
\begin{equation}
H = \sum_j H_\mathcal{S}^{(j)} \otimes \ket{j} \bra{j}_\mathcal{A},
\label{eq:conditional_ham}
\end{equation}
where $\{ \ket{j} \}$ are orthogonal actuator states. The special structure of \eqref{eq:conditional_ham} allows one to evolve $\mathcal{S}$ unitarily by any $H_\mathcal{S}^{(j)}$,
 conditioned on the state of $\mathcal{A}$ \cite{Cappellaro2009, Aiello2015, Hodges2008, Taminiau2014}. This approach---which sidesteps the issue of system-actuator entanglement by working only with states $\{ \ket{j} \}$ of the latter---reduces the problem of indirect quantum control to the well-studied one of direct control; i.e., of synthesising a desired unitary using a set of available system Hamiltonians \cite{Lloyd1995, Alessandro2001, Palao2003, Boscain2006, Chakrabarti2007, Hegerfeldt2013}.

In this paper, we present an explicit scheme for indirect quantum control which employs a resettable actuator to produce unitary evolution of the system, conditioned on the actuator's state. Crucially, our scheme is completely general in that it can be applied for any $\mathcal{S}$ and $\mathcal{A}$. Furthermore, our scheme does not require any particular form of the $\mathcal{S}$-$\mathcal{A}$ Hamiltonian (e.g., we are not restricted to Hamiltonians of the form \eqref{eq:conditional_ham}), but rather, relies on the ability to reset $\mathcal{A}$ rapidly. In this sense, it reduces the general problem of indirect control---for any hybrid system---to the much simpler one of direct quantum control.

\section{The scheme}
\label{sec:scheme}
Consider a quantum system $\mathcal{S}$ coupled to a quantum actuator $\mathcal{A}$. Suppose the pair is initially in the state $\rho(0) = \rho_\mathcal{S} (0) \otimes \rho_\mathcal{A}$, and that $\mathcal{A}$ is periodically reset to $\rho_\mathcal{A}$ at time intervals $\delta t$. (Equivalently, one could couple a succession of fresh actuators to $\mathcal{S}$, each prepared in the state $\rho_\mathcal{A}$.) For convenience, we will assume the resets to be instantaneous, although we shall discuss more realistic resetting later. Between successive resets, $\mathcal{S}$-$\mathcal{A}$ evolves as
\begin{equation}
\dot{\rho}(\tau) = \mathcal{L}(\tau) \rho(\tau),
\end{equation}
where the superoperator $\mathcal{L}(\tau)$ is known as a Liouvillian. We decompose $\mathcal{L}(\tau)$ into parts describing the free dynamics of $\mathcal{S}$ and $\mathcal{A}$, and the coupling between them:
\begin{equation}
\mathcal{L}(\tau) = \mathcal{L}_\mathcal{S} + \mathcal{L}_\mathcal{A} + g(\tau / \delta t) \mathcal{L}_\mathcal{SA}.
\label{eq:H_cycle}
\end{equation}
Here $\mathcal{L}_\mathcal{S}$, $\mathcal{L}_\mathcal{A}$, and $\mathcal{L}_\mathcal{SA}$, which act non-trivially on $\mathcal{S}$, $\mathcal{A}$, and $\mathcal{S}$-$\mathcal{A}$ respectively, are assumed to be time-independent and of Lindblad type \cite{Lindblad1976}. We include an arbitrary piecewise-continuous switching function $g: [0,1] \rightarrow \mathbb{R}$ in \eqref{eq:H_cycle} to demonstrate how time dependence can be incorporated into this scheme. To facilitate comparison with existing indirect control techniques, we will be particularly interested in the case where $\mathcal{S}$-$\mathcal{A}$ evolves unitarily between resets, i.e., where $\mathcal{L}(\tau) \, \cdot = \frac{1}{\irm \hbar} [H(\tau), \, \cdot \,]$, and $H(\tau) = H_\mathcal{S} + H_\mathcal{A} + g(\tau/\delta t) H_\mathcal{SA}$.

We encode the effect on $\mathcal{S}$ of each evolve-and-reset cycle of $\mathcal{A}$ in the dynamical map $\Phi(\delta t)$, which acts on the system's density operator as
\begin{equation}
\Phi(\delta t)  \; \cdot
=
\tr_\mathcal{A}
\left \{
\mathcal{T} \exp \Big[\int_0^{\delta t} \drm \tau \, \mathcal{L}(\tau) \Big]
(\; \cdot \; \otimes \rho_\mathcal{A} )
\right \},
\label{eq:Phi_def}
\end{equation}
where $\mathcal{T}$ denotes the time-ordering operator. We may then write the system's state after the first cycle as $\rho_\mathcal{S}(\delta t) = \Phi(\delta t) \rho_\mathcal{S}(0)$. If $\mathcal{A}$ is reset $n$ times in the interval $[0, t]$ (so that $t = n \, \delta t$, where $n  \in \mathbb{N}$), the reduced state of $\mathcal{S}$ at time $t$ is $\rho_\mathcal{S}(t) = \Phi(t/n)^n \rho_\mathcal{S}(0)$, where $\Phi(t/n)^n$ represents $n$ successive applications of the channel $\Phi(t/n)$.

Our scheme utilizes cycles which are short as compared to the natural dynamics of $\mathcal{S}$-$\mathcal{A}$. We will therefore seek to expand $\rho_\mathcal{S}(t)$ in powers of $1/n$, a small number when the actuator is reset at a high rate. In terms of the dynamical map $ \Phi(t/n)^n$, we wish to find a series of the form
\begin{equation}
\Phi(t/n)^n = 
\Omega_0(t) + \frac{1}{n} \,\Omega_1(t) + 
\frac{1}{n^2} \, \Omega_2(t) + \dots,
\label{eq:Omega_series}
\end{equation}
where each $\Omega_k(t)$ is a superoperator that does not depend on $n$. Noting that $\mathcal{L}( \delta t \cdot \zeta) = \mathcal{L}_\mathcal{S} + \mathcal{L}_\mathcal{A} + g(\zeta) \mathcal{L}_\mathcal{SA}$ is independent of $\delta t$ (since the argument of $g$ is scaled by $\delta t$, see Eq.~\eqref{eq:H_cycle}), we proceed by first expanding $\Phi(\delta t)$ asymptotically for small $\delta t$ as $\Phi(\delta t) = I + \delta t \, \Phi_1 + \delta t^2 \, \Phi_2 + \dots$, where 
\begin{align}
\Phi_k \; \cdot
=
\tr_\mathcal{A}
\int_0^1 \drm \zeta_1 \int_0^{\zeta_1} \drm \zeta_2 \cdots \int_0^{\zeta_{k-1}} \drm \zeta_k \quad
\nonumber \\
\Big[
\prod_{j=1}^k \mathcal{L} (\delta t \cdot \zeta_j)
\Big]
(\; \cdot \; \otimes \rho_\mathcal{A})
\label{eq:Phi_k}
\end{align}
is a superoperator with no $\delta t$ dependence.

Before presenting our main result, let us consider an analogous but simpler situation, namely the case of a matrix $R(\theta)  \, \in \mathsf{SO}(2)$ representing a rotation by $\theta$ in $\mathbb{R}^2$. The action of $R(\theta)$ can be obtained as the outcome of a series of infinitesimal rotations, each given by $I + \delta \theta \, R'(0) + O(\delta \theta^2)$, for an arbitrary $O(\delta \theta^2)$ term. Concretely, setting $\delta \theta = \theta/n$:
\begin{equation}
\lim_{n \rightarrow \infty}
\left[ 
I + \frac{\theta}{n} R'(0) +
O \left( \frac{\theta^2}{n^2} \right)
\right]^n
=
e^{R'(0) \theta} = R(\theta).
\label{eq:analogy}
\end{equation}

The key observation is this: in the limit, only $R'(0)$ contributes while any $O(\delta \theta^2)$ terms, in contrast, are suppressed as $n \rightarrow \infty $ \cite{Gantmacher, Turin}.

\section{Main Result}
Remarkably, the $\Omega_0(t)$ term in Eq.~\eqref{eq:Omega_series} can be evaluated analogously for any system coupled to a quantum actuator that is repeatedly reset. Specifically, Chernoff's theorem  \cite{Chernoff1968} (p.\ 241, see also \cite{Chernoff1970,Chernoff1976}) gives
\begin{equation}
\Omega_0(t) = \lim_{n\rightarrow \infty} \Phi(t/n)^n = e^{ \Phi_1 t }.
\label{eq:chernoff}
\end{equation}
This theorem requires that $\Phi$ be a continuous function of linear contractions on a Banach space, with $\Phi(0) = I$. We verify in the Supplemental Material \footnote{See Supplemental Material at \url{http://link.aps.org/supplemental/10.1103/PhysRevA.93.040301} for additional technical details.} that $\Phi$, as constructed, satisfies these conditions under the induced trace norm  on the set of self-adjoint trace-class operators acting on the system's Hilbert space.

From Eq.~\eqref{eq:Phi_k}, we have that
\begin{equation}
\Phi_1 \; \cdot
=
\mathcal{L}_\mathcal{S} \, \cdot + \bar{g} \tr_\mathcal{A} 
\big[ 
\mathcal{L}_\mathcal{SA}  (\; \cdot \; \otimes \rho_\mathcal{A}) \big],
\label{eq:Phi_1_liouvillian}
\end{equation}
where $\bar{g} \equiv \int_0^1 g(\zeta) \, \drm \zeta$ gives the average coupling strength between $\mathcal{S}$ and $\mathcal{A}$. Observe that to leading order in $1/n$, the system evolves as $\rho_\mathcal{S}(t) = e^{\Phi_1 t} \rho_\mathcal{S}(0)$; therefore, $\Phi_1$ represents an effective Liouvillian for $\mathcal{S}$ in the limit of frequent actuator resets. We now focus on the case where $\mathcal{S}$-$\mathcal{A}$ is closed (i.e., where the bipartite object evolves unitarily between resets of $\mathcal{A}$). Eq.~\eqref{eq:Phi_1_liouvillian} simplifies to
\begin{equation}
\Phi_1 \; \cdot
=
- \frac{\irm}{\hbar} [H_\text{eff}, \, \cdot \;],
\label{eq:Phi_1_simplified}
\end{equation}
where
\begin{equation}
H_\text{eff} = H_\mathcal{S} + 
\bar{g} \tr_\mathcal{A} \big( H_\mathcal{SA} \, \rho_\mathcal{A} \big).
\label{eq:H_eff}
\end{equation}
We show explicitly the details of this calculation in the Supplemental Material \cite{Note1}. Clearly $H_\text{eff}^\dagger = H_\text{eff}$; it follows that, to leading order, the system's reduced dynamics is unitary, and described by
\begin{equation}
\irm \hbar \frac{\drm}{\drm t} \rho_\mathcal{S} (t)
= [H_\text{eff},\; \rho_\mathcal{S}(t)].
\label{eq:von_neumann}
\end{equation}
If, for example, the interaction Hamiltonian has the form $H_\mathcal{SA} = B \otimes C$, the system's reduced dynamics will be well-described by $H_\text{eff} = H_\mathcal{S} + \bar{g} \expval{C} B$, where $\expval{C} = \tr(C \rho_\mathcal{A})$, when the reset rate is high. More generally, different types of coupling between $\mathcal{S}$ and $\mathcal{A}$ will lead to different effective Hamiltonians, generating unitary system dynamics conditioned upon $\rho_\mathcal{A}$. Therefore, by gradually varying the state in which $\mathcal{A}$ is prepared, one can use this scheme to implement entire families of Hamiltonians on $\mathcal{S}$, reducing the problem of indirect control to one of direct quantum control. Moreover, for finite-dimensional systems, our scheme gives universal unitary control of $\mathcal{S}$ provided $H_\mathcal{S}$ and $H_\mathcal{SA}$ are not related by some symmetry \cite{Lloyd1995}.

One can view $\mathcal{S}$ as receiving a discrete kick from $\mathcal{A}$ with each cycle. Notice that $\Phi(\delta t) \rightarrow I$ as $\delta t \rightarrow 0$; in other words, as the reset rate becomes large, the kicks to $\mathcal{S}$ become weaker but more frequent. These competing trends underpin our scheme: When the cycles are short, $\Phi(\delta t) \approx I + \delta t \, \Phi_1$, i.e., the impact of each kick is mostly encoded in $\Phi_1$. As in \eqref{eq:analogy}, the aggregate effect of many such kicks depends only on $\Phi_1$ to leading order. Crucially though, $\Phi_1$ generically describes a non-trivial, but \textit{non-entangling} operation on $\mathcal{S}$-$\mathcal{A}$. This is not unique to our scheme: the non-entangling nature of the first-order Dyson series term has been used in very different contexts, see, e.g., \cite{MM2013}. The present scheme, therefore, exploits the general phenomenon that two quantum objects interacting for a short time $\delta t$ can impact each other faster than they can become significantly entangled. By mimicking a series of short interactions between $\mathcal{S}$ and $\mathcal{A}$ through frequent resets of the latter, we can effectively modify the system's Hamiltonian without significantly increasing the system's entropy. Along the same lines, one can also think of the frequent actuator resets as serving to keep $\mathcal{A}$ in its initial state. The resulting effect is for $\mathcal{S}$ to evolve according to the portion of the total Hamiltonian acting on the system's Hilbert space, up to corrections of order $O(1/n)$.

\section{Example}
To illustrate our scheme, we consider an archetypal problem in quantum control: that of steering a harmonic oscillator $\mathcal{S}$ indirectly via a $d$-level actuator $\mathcal{A}$. Existing techniques depend crucially on $d$ and/or on the nature of the $\mathcal{S}$-$\mathcal{A}$ coupling. For instance, \cite{Vogel1993} requires a Jaynes-Cummings (JC) interaction with $d=2$, \cite{Santos2005} requires a JC-like coupling with $d=3$, and \cite{Jacobs2007} requires switching between two distinct interaction Hamiltonians. Our scheme, in contrast, is model-independent. For illustration, however, we pick $d=2$ and an $\mathcal{S}$-$\mathcal{A}$ Hamiltonian
\begin{equation}
H(\tau) = \hbar \nu a^\dagger a + \frac{\hbar \omega}{2} \sigma_z
+ \hbar g(\tau/\delta t) X \otimes \bold{n} \cdot \boldsymbol{\sigma},
\label{eq:H_example}
\end{equation}
where $X = ( a + a^\dagger)/2$ is a quadrature operator on $\mathcal{S}$ and $\boldsymbol{\sigma} = (\sigma_x, \sigma_y, \sigma_z)$. For different choices of $\bold{n}$, Eq.~\eqref{eq:H_example} describes several hybrid objects amenable to indirect quantum control; for example, a nanomechanical resonator coupled to a superconducting qubit \cite{Irish2003, Jacobs2007, Lahaye2009} or to an electron spin \cite{Rabl2009}, as well as circuit (cavity) quantum electrodynamics setups, where an electromagnetic field mode couples to superconducting qubits (atoms). 

A direct application of Eq.~\eqref{eq:H_eff} yields an effective Hamiltonian
\begin{equation}
H_\text{eff} = \hbar \nu a^\dagger a
+ \hbar \bar{g} \expval{\bold{n} \cdot \boldsymbol{\sigma}}  X,
\label{eq:H_eff_example}
\end{equation}
on $\mathcal{S}$, which is manifestly dependent on $\rho_\mathcal{A}$. It follows from Lloyd's well-known argument in Ref.\ \cite{Lloyd1995} (see also \cite{Lloyd1999}) that any Hamiltonian of the form 
\begin{equation}
H = c_1 a^\dagger a + c_2 X + c_3 P,
\label{eq:algebra}
\end{equation}
can be implemented on $\mathcal{S}$ by suitably varying the state to which $\mathcal{A}$ is reset, where $P = (a-a^\dagger)/2\irm$.

\begin{figure}[h!]
\centering
\includegraphics[width=3.6in]{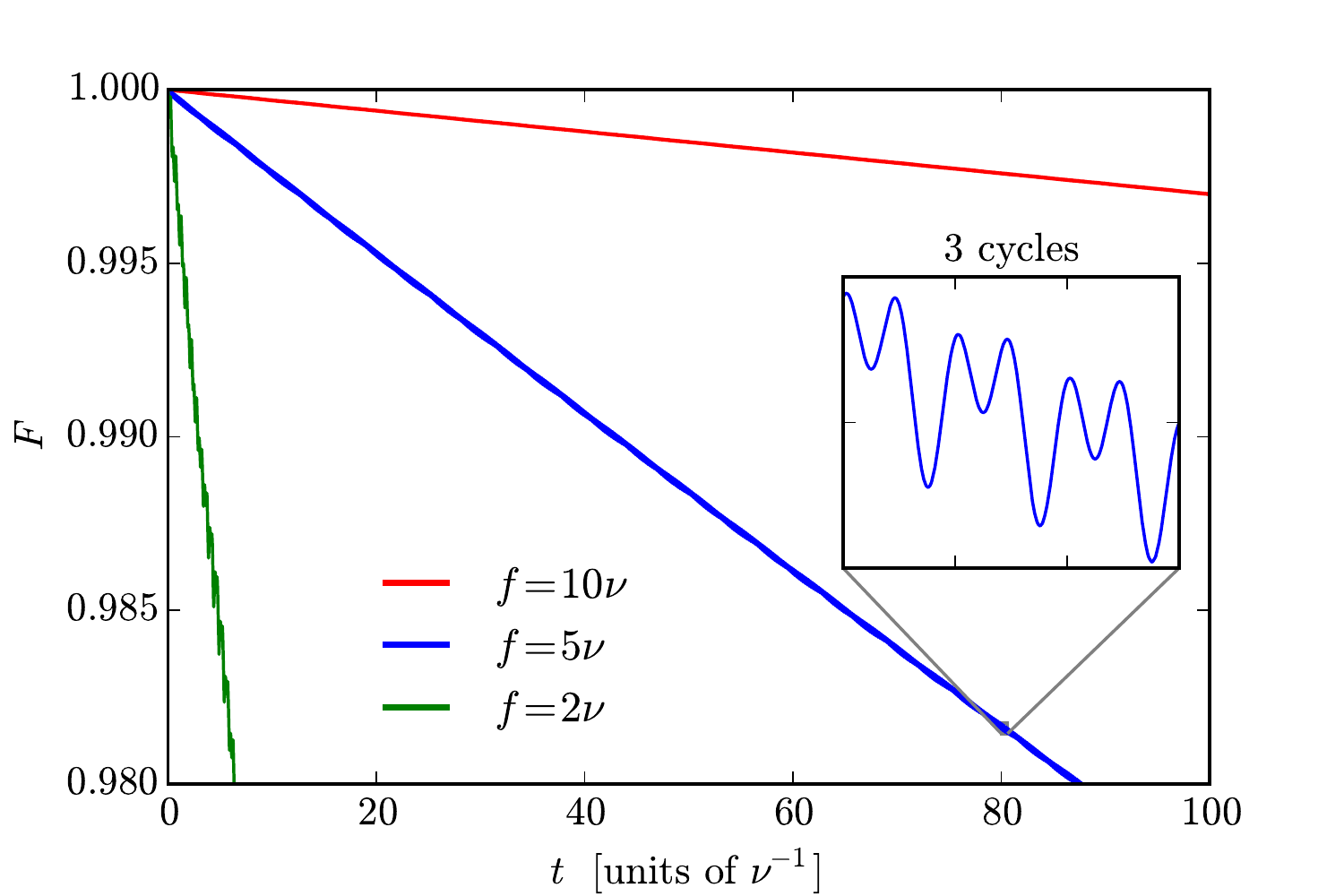}
\caption{(Color online)  Deviation between $H_\text{eff}$ and full $\mathcal{S}$-$\mathcal{A}$ evolution. Shown is the fidelity $F(t) = \sqrt{ \bra{\psi(t)}\,  \rho_\mathcal{S}(t) \, \ket{\psi(t)}}$ between $\ket{\psi(t)} = \exp(- \irm t H_\text{eff} /\hbar) \ket{\alpha}$ and $\rho_\mathcal{S}(t)$, the reduced state of $\mathcal{S}$ under evolution by \eqref{eq:H_example}, for three different resetting rates $f$. (From top to bottom: $f=10\nu$, $f=5\nu$, $f=2\nu$.) The initial state of the system $\rho_\mathcal{S}(0) = \ket{\alpha}\bra{\alpha}$ is coherent with $\alpha = (1 + \irm)/\sqrt{2}$. We set $\nu = \omega $, $\bold{n} = (1,0,0)$ and $\rho_\mathcal{A} = (\openone + \sigma_x)/2$ for illustration, and chose the switching function as $g(\tau/\delta t) = 2\nu\sin^2(\pi \tau/\delta t)$ \cite{Qutip}. }
\label{fig:F_vs_t}
\end{figure}

In implementations, the accuracy with which one can enact evolution by $H_\text{eff}$ through frequent resets of $\mathcal{A}$ will be important. To quantify this accuracy, we consider an initial system state $\rho_\mathcal{S}(0)$ and evolve it numerically for a time $t$ according to the full Hamiltonian \eqref{eq:H_example}, where $\mathcal{A}$ is reset at a rate $f = n/t$. We then compute the fidelity $F$ (see caption of Fig.~\ref{fig:F_vs_t} for definition) as a function of $t$, between the resulting reduced system state and the time-evolved state $e^{- \irm t H_\text{eff}/ \hbar} \rho_\mathcal{S}(0) e^{ \irm t H_\text{eff}/ \hbar}$ that we wish to obtain. The resulting fidelity is shown in Fig.~\ref{fig:F_vs_t} for three different reset rates. 

We observe that two qualitatively distinct types of error arise in implementing $H_\text{eff}$ on $\mathcal{S}$ with our scheme, see Fig.~\ref{fig:F_vs_t}: (i) slowly growing deviation from $F = 1$ as $t$ increases (apparent in the main panel), and (ii) fast ``wiggles" on these otherwise slowly varying curves (shown in the zoomed inset). Type (i) error, which we call \textit{dissipative error}, arises because $\mathcal{S}$ and $\mathcal{A}$ get slightly entangled during cycles of finite duration. Over time, this entanglement causes information about $\mathcal{S}$ to become lost due to the actuator resets, contributing a non-unitary component to system's evolution, which accumulates with every cycle. Thus, dissipative error becomes important on long timescales.

Type (ii) error, which we name \textit{stroboscopic error}, also arises from a finite reset rate $f$; however, it is important only on comparatively short timescales. When $f < \infty$, our scheme approximates smooth evolution by $H_\text{eff}$ using discrete non-unitary kicks from $\mathcal{A}$, which can have complex effects on $\mathcal{S}$. Our asymptotic analysis has been focused on the total effect of each kick, described by Eq.~\eqref{eq:Omega_series} for integer $n$. However, the system can display complicated non-unitary dynamics during a cycle of finite duration, the details of which are not described by $H_\text{eff}$ (nor, more generally, by $\Omega_1(t)$, $\Omega_2(t)$, \dots). Stroboscopic error, then, is the temporary deviation during each cycle between the full open dynamics of $\mathcal{S}$ and the smooth path described in \eqref{eq:Omega_series}. Because this type of error vanishes at the end of every cycle, it does not accumulate with $t$, and hence, is primarily important over short timescales. We will now establish how both the dissipative and the stroboscopic error decay as the cycle length goes to zero. 

\section{Error analysis}
Dissipative error in implementing $H_\text{eff}$ arises from the $\Omega_1(t)$, $\Omega_2(t)$, $\ldots$ terms in \eqref{eq:Omega_series}, which generically introduce dissipation into the system's evolution. These superoperators can be expressed as functions of $\Phi_1$, $\Phi_2$, $\ldots$ by adapting the method developed in \cite{Bentkus2003} (Section 4).  We will use this method to find the form of $\Omega_1(t)$, the leading-order source of dissipative error.

Consider the function 
\begin{equation}
v( \tau) \equiv \Phi( \tau t/n)^n
\,
\Omega_0
\big [ (1-\tau) t \big],
\end{equation}
chosen so that $v(1) = \Phi(t/n)^n$ and $v(0) = \Omega_0(t)$. Observe that with this definition of $v$ we have
\begin{equation}
\Phi(t/n)^n - \Omega_0(t) = v(1) - v(0) = \int_0^1 v'(\tau) \, \drm \tau.
\end{equation}
One rapidly arrives at expressions for $\Omega_1(t)$, $\Omega_2(t)$, $\ldots$ by expanding $v'(\tau)$ asymptotically in $1/n$ and matching powers with Eq.~\eqref{eq:Omega_series}. In particular:
\begin{equation}
\frac{1}{n} \, \Omega_1(t)
= \frac{t^2}{n} \int_0^1
e^{\Phi_1 \tau t}
\Big( \Phi_2 - \frac{1}{2}\Phi_1^2 \Big)
e^{\Phi_1 (1-\tau) t} \, \drm \tau.
\label{eq:Omega_1}
\end{equation}
In terms of the reset rate $f = n/t$, Eq.~\eqref{eq:Omega_1} scales as $O(t/f)$. (This statement is readily formalized by noting that the induced trace norm of the integral in \eqref{eq:Omega_1} is independent of $t$.) The $t/f$ scaling is apparent in Fig.~\ref{fig:F_vs_t}, where the reset rates considered are sufficiently high that $\Omega_1(t)$ is the dominant source of dissipative error. In particular, observe that for each $f$ plotted, the deviation between the reduced dynamics and the  $H_\text{eff}$-generated trajectory (corresponding to $\Omega_0(t)$, the leading-order term in $1/n$) is nearly linear in $t$. Furthermore, the slope of the lines scale inversely with $f$.

We now turn our attention to the stroboscopic error. The dynamical map in Eq.~\eqref{eq:Omega_series} describes the system's evolution for a time $t = n \, \delta t$, corresponding to an integer number of cycles. In other words, it gives $\rho_\mathcal{S}$ at the end of each cycle. However, between successive resets (i.e., mid-kick), $\mathcal{S}$ will temporarily stray from the smooth trajectory given in \eqref{eq:Omega_series}.

One can quantify the stroboscopic error by comparing the desired evolution by $H_\text{eff}$ with the system's reduced dynamics between successive actuator resets. We use a truncated Dyson series to perform this comparison: If $\rho_\mathcal{S}(t_n)$ is the system's state after cycle $n$, then during cycle $n+1$, evolution by $H_\text{eff}$ would give
\begin{equation}
\rho_\mathcal{S}^{(\text{eff})}(t_n + \tau) 
\stackrel{O(\tau)}{=}
\rho_\mathcal{S}(t_n) - \frac{\irm \tau}{\hbar}
[H_\text{eff},\; \rho_\mathcal{S}(t_n)],
\label{eq:stroboscopic_eff}
\end{equation}
for $0 \le \tau \le \delta t$. When the cycles are short, i.e., when $\tau \le \delta t \ll 1$ in units set by the largest characteristic frequency of $H(\tau)$, the $O(\tau^2)$ terms in the Dyson series will be subdominant, and so we work only to first order in $\tau$.

Eq.~\eqref{eq:stroboscopic_eff} is the evolution our scheme seeks to implement. However, the actual full evolution of $\mathcal{S}$ between successive resets is computed by evolving $\mathcal{S}$-$\mathcal{A}$ according to $H(\tau)$ and then tracing out the actuator:
\begin{equation}
\rho_\mathcal{S}^{(\text{full})}(t_n + \tau)
\stackrel{O(\tau)}{=}
\rho_\mathcal{S}(t_n) - \frac{\irm}{\hbar}
\tr_\mathcal{A}
\big[ \! \int_0^\tau \!\! H(\tau') \drm  \tau', \; \rho_\mathcal{S}(t_n) \, \otimes \, \rho_\mathcal{A} \big].
\label{eq:stroboscopic_full}
\end{equation}
Concretely, stroboscopic error is described by the difference between Eqs.~\eqref{eq:stroboscopic_eff} and \eqref{eq:stroboscopic_full}:
\begin{flalign} \label{eq:stroboscopic_diff}
&\rho_\mathcal{S}^{(\text{eff})}(t_n + \tau)
-
\rho_\mathcal{S}^{(\text{full})}(t_n + \tau)
\stackrel{O(\tau)}{=} &  \\
& \quad \;
- \frac{\irm \tau}{\hbar} 
\left \{
\bar{g} - \frac{\delta t}{\tau} \int_0^{\tau/\delta t}
\! \! g(\zeta) \, \drm \zeta
\right \}
\big[ \tr_\mathcal{A}(H_\mathcal{SA} \, \rho_\mathcal{A}), \; \rho_\mathcal{S}(t_n) \big]. \nonumber
\end{flalign}
The braced term in \eqref{eq:stroboscopic_diff} is bounded above in magnitude by $2 g_\text{max} (\delta t - \tau)/ \tau$, where $g_\text{max}$ is the largest coupling strength attained in each cycle. (Thus, stroboscopic error is reduced when the $\mathcal{S}$-$\mathcal{A}$ coupling remains weak.) The commutator, in contrast, can vary arbitrarily with $t_n$, depending on the nature of $H(\tau)$. However, it is always suppressed by a prefactor of $\delta t - \tau$, and so  Eq.~\eqref{eq:stroboscopic_diff} generically scales as $O( \delta t - \tau)$, which is upper-bounded by $O(\delta t)$. In terms of reset frequency then, stroboscopic error scales as $O(1/f)$.

Unlike dissipative error, which accumulates with $t$, stroboscopic error vanishes with each reset. Thus, while the former type is reduced by implementing $H_\text{eff}$ for short durations $t$, the latter can be sidestepped entirely by choosing $f$ and $t$ to give an integer number of cycles. In general, both types of error can be suppressed arbitrarily by choosing an $f$ that is large compared to characteristic frequencies of $\mathcal{S}$ and $\mathcal{A}$.

\section{Discussion \& Outlook}
Our scheme is somewhat reminiscent of the quantum Zeno effect (QZE) \cite{Misra1977, Facchi2002, Layden2015}; both our scheme and the QZE feature emergent unitary dynamics generated by an effective Hamiltonian, which results from rapidly repeated operations. However, our scheme does not invoke any type of measurement, nor any notion of ``leakage" between measurement eigenstates. Moreover, the QZE typically involves repeated kicks which remain strong (or at least non-vanishing) in the high-frequency limit \cite{facchi2008}. In contrast, the effect on $\mathcal{S}$ of each individual cycle in our scheme vanishes as $f \rightarrow \infty$. Zanardi and Campos Venuti recently discovered a phenomenon closely related to the QZE, wherein unitary evolution, generated by a ``dissipation-projected Hamiltonian", can arise in an open system. While their results and ours may both stem from a common fundamental principle, the scaling that underpins Refs.\ \cite{Zanardi2014,Zanardi2015} is entirely absent in our scheme.

While we have assumed for illustration that $\mathcal{A}$ is reset instantaneously and at regular intervals, neither of these idealizations are essential to our scheme. Notice from Eq.~\eqref{eq:Phi_1_liouvillian} that $\mathcal{L}_\mathcal{A}$ does not contribute to $\Phi_1$; therefore, $\mathcal{S}$ would still evolve unitarily by $H_\text{eff}$ to leading order if $\mathcal{A}$ were reset through a gradual non-unitary process. We note also that Chernoff's theorem---which gives the evolution of $\mathcal{S}$ in the high $f$ regime---can be generalized to describe cycles of non-uniform duration. Specifically, when $n$ actuator resets are performed in a time $t$, the system's evolution will be well-described by $H_\text{eff}$ (or more generally, by $\Phi_1$) when the longest time between actuator resets is sufficiently short \cite{Smolyanov1999, Smolyanov2003}.

Finally, we wish to compare our control scheme with the one proposed in \cite{Lloyd2001}: Whereas we employ a resettable quantum actuator to achieve indirect unitary control of a system, Lloyd and Viola used a resettable and \textit{measurable} ancilla to achieve arbitrary open-system dynamics. The similar requirements of both schemes suggest the possibility of implementing arbitrary open-system dynamics on $\mathcal{S}$ through indirect control.



\section{Acknowledgements}
AK, EMM and DL acknowledge support from the NSERC Discovery and NSERC PGSM programs. DL also acknowledges support through the Ontario Graduate Scholarhip (OGS) program. The authors thank Lorenza Viola, Paola Cappellaro, and Daniel Grimmer for helpful discussions.

\bibliography{references}

\end{document}